\documentclass[aps,prd,amsmath,amssymb,showpacs,showkeys]{revtex4}
\usepackage{graphicx}
\usepackage{bm}
\newcommand{\be}{\begin{equation}}
\newcommand{\ee}{\end{equation}}
\newcommand{\bea}{\begin{eqnarray}}
\newcommand{\eea}{\end{eqnarray}}
\newcommand{\nn}{\nonumber\\}
\def\journal#1#2#3#4{{#1} {\bf #2}, #3 (#4)}
\def\eq#1{(\ref{#1})}
\def\hf{\frac{1}{2}}
\def\v#1{\mathbf{#1}}
\def\mr#1{\mathrm{#1}}

\def\tr{\mr{tr}}

\def\la{\langle}
\def\ra{\rangle}

\def\phid{\phi^\dagger}

\begin{document}
\title{Boundary conditions and consistency of effective theories}
\author{Janos Polonyi$^a$, Alicja Siwek$^{ab}$}
\affiliation{$^a$University of Strasbourg,
High Energy Physics Theory Group, CNRS-IPHC,
23 rue du Loess, BP28 67037 Strasbourg Cedex 2, France}
\affiliation{$^b$Wroclaw University of Technology, Institute of Physics, Wybrzeze~Wyspianskiego~27, 50-370 Wroclaw, Poland}
\begin{abstract}
Effective theories are non-local at the scale of the eliminated heavy particles 
modes. The gradient expansion which represents such non-locality must be truncated 
to have treatable models. This step leads to the proliferation of the degrees
of freedom which renders the identification of the states of the effective theory
nontrivial. Furthermore it generates non-definite metric in the Fock space 
which in turn endangers the unitarity of the effective theory. It is shown that
imposing a generalized KMS boundary conditions for the new degrees of
freedom leads to reflection positivity for a wide class of Euclidean effective theories,
thereby these lead to acceptable theories when extended to real time.

\end{abstract}
\date{\today}
\pacs{}
\keywords{Effective theories, reflection positivity, indefinite norm, unitarity}
\maketitle

\section{Introduction}
The observed richness of the scale dependence of fundamental physical laws
renders the project of constructing the ultimate Theory of Everything
unpractical. Instead, effective theories are put forward, models with
limited range of applicability, as a less ambitious but more realistic 
alternative. Such a model is derived from an underlying, more microscopic
theory by eliminating the degrees of freedom which belong to short
distances not resolved by the effective model.

The elimination of a propagating particle mode from a local theory induces
long range correlations in the remaining effective dynamics. It is easy
to verify in the framework of imaginary time, Euclidean quantum field theories
that these non-local features arising from the elimination of massive 
particle modes can be retained as higher order derivative terms 
in the effective action. By restricting our interest to sufficiently low 
energy the gradient expansion can be truncated. We consider in this work two
questions raised by this step, the specification of the states and the consistency
of the effective theory.

Let us consider a single degree of freedom which is governed by a Lagrangian containing
time derivatives up to $n_d$-th order. Its classical phase space is $2n_d$ dimensional. 
Assuming that ordinary, covariant particles are defined by the usual, velocity dependent
Lagrangians we may match this space by the phase space of $n_d$ ordinary particles. 
One arrives at similar conclusions in quantum physics when time derivatives up to order
$n_d$ are added to the Schr\"ordinger equation and one introduces $n_d$ component
wavefunctions, consisting of the first $n_d-1$ time derivatives of the original wavefunction. 
One may gain a new insight into the appearance of antiparticles in relativistic
quantum mechanics in this manner. The issue of identifying the states of an
effective theory with higher order derivatives is therefore nontrivial. This complication
is actually natural because in order to arrive at a well defined effective theory we
have to specify the initial conditions for the heavy particles which are eliminated.
The specification of the initial or final states amounts to the definition of boundary
conditions in time within the path integral formalism.
Notice that this problem is bound entirely to the truncation of the effective dynamics. 
In fact, when the heavy particles with well defined initial conditions are eliminated 
exactly then the resulting integro-differential equations yield unique solutions when
the usual, $n_d=1$ initial conditions are imposed on the light particles. But we
have to keep in mind that no exact equations of motion are known in physics, we should 
allow the presence of higher order time derivatives in the Lagrangian with some small 
coefficients in all theories.

The other question related to the truncation is that even if the underlying theory is 
consistent, follows a unitary time evolution, the effective model, defined by a 
truncated gradient expansion may show inconsistencies and provide non-unitary time
evolution for the light particles. Consider for instance a solid state at low enough 
temperature, described in terms of electrons and phonons. The periodic ion core potential
induces a band structure which can be modelled by a Lagrangian for the electrons without
potential but containing higher order derivatives in the kinetic energy. Such a theory 
becomes unusable at sufficiently high energy where lattice defects may occur. The true
physics requires new degrees of freedom which should appear as a loss of unitarity
of the simple effective model based on electrons and phonons. We argue below that these 
two problems are related, an appropriately chosen subspace of the states of the effective
theory leads to consistent dynamics in the natural manner.

To see our problems in a simpler setting consider a scalar theory defined by the action
\be
S[\phi]=\int dx\left[\phi\left(\sum_{n=1}^{n_d}c_n\Box^n\right)\phi(x)-V(\phi(x))\right]
\ee
with derivatives up to order $n_d$. We assume that the model obeys time reversal 
invariance, the coefficients $c_n$ and the potential $V(\phi)$ are real, furthermore
$(-1)^{n_d}c_{n_d}>0$. For the sake of simplicity we take $n_d$ an odd integer.
The generator functional for the Green functions can be written as
\be\label{genfunct}
\int D[\phi]e^{iS[\phi]+i\int dxj(x)\phi(x)}=e^{-i\int dxV(\frac{\delta}{i\delta j(x)})}e^{-\frac{i}2\int dxdyj(x)D(x-y)j(y)}
\ee
where the free propagator is
\be
D(p)=\left(\sum_{n=1}^{n_d}(-1)^nc_n(p^2)^n\right)^{-1}
\ee
in the momentum-space, $i\epsilon$ prescription being suppressed. Its partial 
fraction decomposition,
\be\label{partfrac}
D(p)=\sum_{j=1}^{n_d}\frac{Z_j}{p^2-m^2_j}
\ee
clearly involves negative contributions, there is at least one $Z$ factor which is 
negative. The free generator functional, the second exponential factor in Eq. \eq{genfunct},
suggests the presence of states with negative norm and may involve complex energy
states $m^2_j<0$, when the decomposition \eq{partfrac}
is used \cite{nagy}. Though the effective model which retains the dynamics of the eliminated particle
modes exactly remains consistent the truncation of the gradient expansion may 
introduce instability and loss of unitarity. In fact, some particles
introduced by the decomposition \eq{partfrac} may have complex energy, $m_j^2<0$,
and it is not clear if the time evolution remains unitary within the subspace of 
physical, positive norm states. These questions raise concern about the feasibility of 
the effective theory strategy. We shall see that the negative norm states provide
a common framework to address both issues mentioned above, namely the specification of
states and the consistency.

The higher order derivative terms have already been considered in quantum field theory
as regulators \cite{dirac,podolski,pauli,schwed,villars,feynman} and they are 
known to generate states with negative norm \cite{pais}. An extension of the Higgs sector of the
Standard Model involving higher order derivative terms in the kinetic energy has
been proposed \cite{jansen}, too. Negative norm states occur in the covariant
quantization of gauge theories as well \cite{bleuer,gupta} but they can safely be
excluded from the asymptotic states by means of gauge symmetry without upsetting 
the unitary time evolution in the physical, positive norm sector. Without such a powerful symmetry argument the 
fate of the instability and unitarity becomes a more difficult problem
for the effective models to resolve. 

It was conjectured that instability and loss of unitarity, generated by ghost 
particles with complex energy and negative norm can be excluded from the asymptotic 
states of finite energy because the mass of a ghost particle diverges with the 
cut-off $\Lambda$ \cite{leewick}. The role of the cut-off is taken over the heavy 
particle energy scale in the context of the effective models. This argument 
eliminates safely the states with negative norm but leaves a delicate point
to settle, the case of zero norm states. These states which arise in a natural
manner in a linear space with indefinite metric may develop exponential growth 
in time in their amplitude and spoil the stability of the model. The proposal,
put forward in Refs. \cite{leewick} is to impose boundary conditions which eliminate
the unstable modes. Similar treatment has been pursued in excluding the runaway
solution of the radiation reaction problem in classical electrodynamics \cite{diracr}.
Such boundary conditions imposed in the future appear rather ad hoc
and it generates acausal effects. To make things worse, the absence of the 
exponentially growing amplitudes eliminated by such boundary conditions leads to
non-unitary time evolution. The perturbative procedure to restore unitarity
by the appropriate modification of the imaginary part of the Feynman graphs
runs into difficulties \cite{boulware}.

It will be shown in a non-perturbative manner, by means of lattice regularization
in time that the method of classical field theory to treat the higher order time derivatives
can be carried over to the quantum case. Furthermore, it is found that the coordinates 
corresponding to even or odd order time derivatives are represented by self- or 
skew-adjoint operators, both satisfying the standard canonical commutation relations. 
An important implication of this correspondence is that the positive 
definite (indefinite) space belongs to self(skew)-adjoint operators and the trajectories in the path integral
are periodic (antiperiodic) in time when expectation values are calculated.
This structure which is reminiscent of the KMS construction is used to establish consistency
in a non-perturbative manner. The key is the demonstration of
reflection positivity for the truncated theories in imaginary time. This argument 
will be demonstrated in the case of an effective Yang-Mills-Higgs models. 

Section \ref{indnorm} is a brief
summary of the salient features of quantum mechanics on linear space with 
indefinite norm. A generic effective theory, the Yang-Mills-Higgs model with 
higher order derivatives is introduced in Section \ref{effmod}. The model is considered in 
lattice regularization in Euclidean space-time and reflection positivity
is demonstrated in the Fock space span by local operators with positive
time reversal parity. This property is sufficient to assure the Wightman axioms 
for the real-time theory \cite{osterwalder}. Finally, the summary of our 
results is given in Section \ref{sum}.

\section{Quantum mechanics}\label{indnorm}
Quantum mechanics constructed on linear spaces with indefinite norm 
\cite{nagy,leewick,boulware} has a number of unusual features. These are summarized
below by paying special attention to the relation of the signature of the norm
with the definition of the adjoint of an operator, the closing relations,
the canonical commutation relations and the path integral expressions. The obvious
motivation of reviewing these properties is to develop simple means to recognize the
restrictions of the linear space and operators which lead to a positive 
definite linear space and physically interpretable structure. The states with 
negative or vanishing norm appear at intermediate time only, the physical
asymptotic states must have positive norm.

The linear space $H$ with positive definite metric is defined by means of the scalar
product $\la u|v\ra$ satisfying the requirements (i) $\la u|v\ra=\la v|u\ra^*$, 
(ii) $\la u|(a|v\ra+b|w\ra)=a\la u|v\ra+b\la u|w\ra$,
(iii) $\la u|u\ra\ge0$, and (iv) $\la u|u\ra=0$ for $|u\ra=0$ only.
We shall use decomposable spaces with non-definite metric where 
properties (iii) and (iv) are replaced by the conditions (iii') $H=H_++H_-$ 
where $H_\pm=\{|u\ra|\la u|u\ra\gtrless0\}$ 
with $\la H_+|H_-\ra=0$, and (iv') each vector $|u\ra$ can be written
as $|u\ra=|u_+\ra+|u_-\ra$, $\la u_\pm|x_\pm\ra\gtrless0$ in a unique manner \cite{nagy}.
Note that decomposability, (iv'), makes the metric non-degenerate by excluding
zero norm states orthogonal to the rest of the space.
We assume furthermore that our linear space can be upgraded to a Hilbert space 
and can be made complete with respect to the scalar product $\la u|v\ra'=\la u_+|v_+\ra-\la u_-|v_-\ra$.

The classification of the operators is a more involved question now because it is
based on two non-trivial quadratic forms, the one given by the operators and 
another one, the scalar product. Let us assume that our linear space is separable
and use a basis $\{|n\ra\}$ where the non-definite metric $\eta_{mn}=\la m|n\ra$ 
is a non-degenerate Hermitian matrix, $\eta^\dagger=\eta$, which can be brought
into the normalized diagonal form $\eta_{mn}=\pm\delta_{m,n}$ by the choice of an appropriate basis.
The matrix elements $A_{jk}$ of an operator $A$ are defined in this basis by the equation
\be
\la m|A|n\ra=\sum_k\eta_{mk}A_{kn}.
\ee
The non-trivial metric makes the adjoint and the Hermitian adjoint of an operator different.
The adjoint $\bar A$ of an operator $A$ is defined by $\la u|\bar A|v\ra=\la v|A|u\ra^*$
what gives $\bar A=\eta^{-1}A^\dagger\eta$. The self- or skew-adjoint operators 
satisfy the condition $\bar A=\sigma_AA$ with the sign $\sigma_A=+1$ and $-1$, respectively. 
Two eigenvectors, $A|\lambda\ra=\lambda|\lambda\ra$, $A|\rho\ra=\rho|\rho\ra$ give
\be\label{eigval}
(\lambda-\sigma_A\rho^*)\la\rho|\lambda\ra=0.
\ee	
Thus the spectrum is real or imaginary for self- or skew-adjoint operators, respectively
in the subspace of orthogonal eigenvectors with non-vanishing norm.
Furthermore, two eigenstates $|\lambda\ra$ and $|\rho\ra$ can have non-vanishing 
overlap only if their eigenvalues are related, $\lambda=\sigma_A\rho^*$, allowing 
real spectrum for skew-adjoint operators with non-orthogonal eigenvectors.

Let us now consider a free particle whose dynamics is based on the canonical pair 
of operators $\hat q_\sigma$ and $\hat p_\sigma$ satisfying $[\hat q_\sigma,\hat p_\sigma]=i$.
These operators are either self- or skew-adjoint and by insisting on real spectrum 
we have to give up the orthogonality of the eigenstates in the skew-adjoint case 
and have to use $\eta(q,q')=\delta(q-\sigma q')$ according to Eq. \eq{eigval}.
We shall need the closing relation in coordinate basis
\be\label{clc}
\openone=\int dq|\sigma q\ra\la q|.
\ee
The equations $e^{i\hat pq'}\hat qe^{-i\hat pq'}=\hat q+q'$ 
and $\la q|p\ra=e^{ipq}/\sqrt{2\pi}$ yield
\be
\eta(p,p')=\la p|p'\ra=\int\frac{dq}{2\pi}e^{-iq(p-\sigma p')}=\delta(p-\sigma p')
\ee
and the closing relation
\be\label{cli}
\openone=\int dp|\sigma p\ra\la p|
\ee
in momentum space. The self-adjoint operators obviously realize linear space with
definite norm. The Pauli matrix $\sigma_x$ has eigenvalues $\pm1$, showing
that the skew-adjoint case leads to indefinite norm.

Anticipating applications in quantum field theory let us consider a harmonic oscillator
of Hamiltonian 
\be\label{hoham}
\hat H_\sigma=\frac\sigma2(\hat p^2_\sigma+\hat q^2_\sigma)=\sigma\bar a_\sigma a_\sigma,
\ee
given in terms of the operator $a_\sigma=(\hat q_\sigma+i\hat p_\sigma)/\sqrt{2}$ satisfying the commutation relation
$[a_\sigma,\bar a_\sigma]=\sigma$. One can easily construct the Hilbert space
where these operators act in an irreducible manner. It is enough to treat the case
$\sigma=+1$, the other linear space will be given by the exchange $a\leftrightarrow\bar a$.
We take the operators $b=a_+$, $\bar b=\bar a_+$ and start with the usual assumption, the 
existence of an eigenstate of the self-adjoint operator $\bar bb$, 
$\bar bb|\lambda\ra=\lambda|\lambda\ra$ and consider the double infinite series of states 
$\cdots,b^2|\lambda\ra,b|\lambda\ra,|\lambda\ra,\bar b|\lambda\ra,\bar b^2|\lambda\ra,\cdots$
corresponding to the eigenvalues $\cdots,\lambda-2,\lambda-1,\lambda,\lambda+1,\lambda+2,\cdots$
of $\bar bb$.  If $\lambda$ is non-integer then this series is infinite on both 
ends and the Hamiltonian is unbounded. Bounded Hamiltonian requires that the
series stops, either to the left or to the right. The equations $\la\lambda|\bar bb|\lambda\ra=\lambda\la\lambda|\lambda\ra$
and $\la\lambda|b\bar b|\lambda\ra=(\lambda+1)\la\lambda|\lambda\ra$ require
$\lambda$ to be integer and the stopping at the left or the right end corresponds 
to the series $\lambda\ge0$ or $\lambda\le-1$, respectively. In the former case
we find definite norm, $\mr{sign}(\la\lambda+1|\lambda+1\ra)=\mr{sign}(\la\lambda|\lambda\ra)$
and the latter implies norm with both signs, $\mr{sign}(\la\lambda-1|\lambda-1\ra)=-\mr{sign}(\la\lambda|\lambda\ra)$.
Therefore the canonical operator algebra realized with self- or skew-adjoint operators
corresponds to definite or indefinite norm, respectively and the Hamiltonian \eq{hoham} possesses a stable ground state.

We may find yet another characterization of the signature of the metric in the
linear space of states. The time reversal $\Theta$ is an anti-unitary transformation,
acting in the Schr\"odinger representation as $\Theta:A\to\bar A$ on the operators
and one may consider operators with well defined time reversal parity 
$\Theta A=\tau_A A$, with $\tau_A=\pm1$. In the Heisenberg representation the 
transformation properties of the Heisenberg
equation of motion under time reversal requires the slight extension
$\Theta:A(t)\to\bar A(-t)$, in particular the time derivative contributes to the
time inversion parity by $-1$, $\tau_{\partial_0A}=-\tau_A$. The Heisenberg
commutation relations require that the canonically conjugated pairs share the same
time reversal parity. The time parity of the canonical operators $\tau_q$ determines
the signature of the linear space because $\tau_A=\sigma_A$.

Finally, we consider the path integral representation of the time evolution amplitude 
for a system with a self-adjoint $\hat q$ and a skew-adjoint $\hat q'$ coordinate,
\be
\la q_f,-q'_f|e^{-itH}|q_i,q'_i\ra
=\int D[p]D[p']D[q]D[q']e^{i\int dt[p\dot q+p'\dot q'-H(q,q',p,p')]},
\ee
where $H(q,q',p,p')=\la q,q'|\hat H|p,p'\ra/\la q,q'|p,p'\ra$
is a complex function for self-adjoint Hamilton operator $\hat H$, 
$H^*(q,q',p,p')=\la p,-p'|H|q,q'\ra/\la p,-p'|q,q'\ra=H(q,-q',p,-p')$. 
We regain real $H(q,q',p,p')$ and the phase space path integral can be rendered 
well defined by the usual $i\epsilon$ prescription for time reversal invariant 
dynamics. The integration over the momentum trajectories can easily be carried 
out for the Hamiltonian of the type $\hat H=(\hat p^2-\hat p'^2)/2+U(\hat q,\hat q')$.
Note the unusual sign in the kinetic energy of the degree of freedom represented by 
with the skew-adjoint operators, as a result of the coefficient $\sigma$ in the Hamiltonian
\eq{hoham}. As a result, the path integral in the coordinate space is
\be\label{cpathint}
\la q_f,-q'_f|e^{-itH}|q_i,q'_i\ra
=\int D[q]D[q']e^{i\int dt[\hf\dot q^2-\hf\dot q'^2-U(q,q')]}
\ee
and it displays an unusual sign in the kinetic energy of this degree of freedom.

\section{Reflection positivity}\label{effmod}
We address now the conditions of arriving at consistent dynamics in a model
with higher order time derivatives.  We do it in the context of the Yang-Mills-Higgs 
model for imaginary time with a scalar matter field whose kinetic energy contains 
higher order derivatives. The action is written as
\be\label{action}
S[\phi,\phid,A]=\int d^dx\left[K(D)-\phid L(D^2)D^2\phi+V(\phid\phi)\right],
\ee
where the gauge field is given as $A_\mu=A_\mu^a\tau^a$, $\tau^a$ denoting the
generators of the gauge group, $D_\mu=\partial_\mu-iA_\mu$ stands for the 
covariant derivative and $\Lambda$ is the UV cut-off. The functions $K$ 
and $L$ are bounded from below and are chosen to be a polynomial of the
covariant derivative of order at most $n_d$ and $n_d-2$, respectively. 
For instance, the perturbatively renormalizable theory is realized by the
choice $K(D)=-\tr([D_\mu,D_\nu])^2/2g^2$ and $L=1$. 

The canonical treatment of a classical dynamical systems with higher order time derivatives has
been worked out for a long time \cite{ostrogadksi}, it is based on the introduction 
of a new coordinate, together with its canonical pair for each higher order 
derivative except for the last one, $A_{j\mu}(x)=D_0^jA_\mu(x)$ and
$\phi_j(x)=\partial_0^j\phi(x)$ for $j=0,\ldots,n_d-1$ in our case. It will be shown
below that this apprach is appropriate for quantum fields as well by explicitly 
constructing the path integral for the fields $A_{j\mu}(x)$ and $\phi_j(x)$. 

The higher derivative terms in the action \eq{action} indicate the presence of states with 
negative norm in the Fock space. The time reversal parity of the coordinates 
$A_{j\mu}(x)$ and $\phi_j(x)$ is $\tau=(-1)^{j+\delta_{\mu,0}}$ and $(-1)^j$,
respectively. The states ${\cal F}|0\ra$ where ${\cal F}$ is a local gauge invariant 
functional of the fields of even time reversal parity should span a Fock space of 
definite norm. This argument is naturally formal because it is just the difficulty
with the definiteness of the norm which prevents us from establishing the quantum 
counterpart of this classical approach in a reliable manner. Nevertheless
its conclusion remains valid as will be shown below. In what follows we
use the action \eq{action} directly in the path integral formalism where it is 
usually derived instead of extending the quantization procedure for higher order 
derivatives.

The strategy to give meaning to the theory \eq{action} is to find a set of
fields whose Green functions satisfy Wightman's axioms when continued to real
time. The set of conditions which is necessary and sufficient for this to 
happen has been found \cite{osterwalder} and its piece which requires explicit
justification is reflection positivity. It is usually tested in lattice regularization
where it appears as the positivity of the transfer matrix in imaginary time.
Therefore we consider the model for imaginary time where lattice regularization 
is imposed. What is crucial is the introduction of a finite step size in time,
the spatial directions will be discretized only because there is no other way
to regulate gauge theories in a non-perturbative manner. 
The lattice fields are $\phi(n)=a\phi(x)$, $\phid(n)=a\phid(x)$, 
$U_\mu(n)=U^\dagger_{-\mu}(n+\hat\mu)=e^{igaA_\mu(n)}$, with $a$ as lattice spacing
and $\hat\mu$ denotes the unit vector of direction $\mu$.
The gauge transformation is represented by $\phi(n)\to\omega(n)\psi(n)$,
$\phid(n)\to\omega^\dagger(n)\psi^\dagger(n)$ and
$U_\mu(n)\to\omega(n+\hat\mu)U_\mu(n)\omega^\dagger(n)$ and the covariant derivative is 
replaced by the finite difference $D_\mu\phi(n)=U^\dagger_\mu(n)\phi(n+\hat\mu)-\phi(n)$,
in particular $D^2\phi(n)=\sum_\mu[U^\dagger_\mu(n)\phi(n+\hat\mu)+U_\mu(n-\hat\mu)\phi(n-\hat\mu)-2\phi(n)]$.
The partition function of the bare Euclidean theory is the path integral
\be\label{latpint}
Z=\int D[U]D[\phid]D[\phi]e^{-S_L}
\ee
with the lattice action
\be\label{laction}
S_L=\sum_n\sum_{\gamma'} a_{\gamma'}\tr U_{\gamma'}(n)
+\sum_n\phid(n)\sum_\gamma U^\dagger_\gamma(n)\phi(n+\gamma)b_\gamma+\sum_nV(\phid(n)\phi(n))
\ee
where $\gamma'$ and $\gamma$ denote closed and open paths respectively, up to 
length $n_d$, $n+\gamma$ stands for the lattice site where the path $\gamma$
arrives upon starting at $n$ and $U_\gamma(n)$ is the path ordered product of the
link variables along this path. Time reversal invariance requires that for each
path $\gamma$ its time reversed version $\Theta\gamma$  be included in the sums
with $a_{\Theta\gamma'}=a^*_{\gamma'}$, $b_{\Theta\gamma}=b^*_\gamma$, making the action $S_L$ real.

Guided by the classical procedure \cite{ostrogadksi} we construct $n_d$ lattice 
field variables in the following manner \cite{defour}. We pick $n^0$ as the time coordinate on the
lattice and regroup each $n_d$ consecutive space-like hyper-surface into a single
blocked time slices of the Euclidean space-time. The new fields, defined as functions of 
the new time variable are
\be
\phi_j(t,\v{n})=\phi(n_dt+j,\v{n}),
\ee
and
\be
U_{j,\mu}(t,\v{n})=U_\mu(n_dt+j,\v{n})
\ee
where $t$ is integer and $j=1,\cdots,n_d$ and we write the action \eq{laction} as
\be\label{blaction}
S_L=\sum_t[L_s(t)+L_{km}(t)+L_{kg}(t)]
\ee
where
\bea
L_s(t)&=&S_s[U(t),\phid(t),\phi(t)],\nn
L_{kg}(t)&=&S_{kg}[U(t),U(t+1)],\nn
L_{km}(t)&=&\sum\limits_{t,\v{m},\v{n}}\phid_j(t+1,\v{m})
\Delta_{j,k}(\v{m},\v{n};U(t),U(t+1))\phi_k(t,\v{n})+c.c.
\eea
Here $S_s[U,\phid,\phi]$ includes field variables within a single blocked time slice,
the expression $S_{kg}[U,U']$ collects the contributions to the gauge field action
which contain the product of link variables $U$ and $U'$ belonging to two 
consecutive blocked time slices. The matter inter-block kinetic term $L_{km}$
represents the coupling of the matter field located at consecutive blocked time 
slices, containing the link variables $U$ and $U'$ of these two time slices and 
finally, c.c. stands for complex conjugate. One introduces the transfer matrix $T$ defined by
\be
\la\phid,\phi,U|T|\phi'^\dagger,\phi',U'\ra=e^{L_{kg}[U,U']
+\hf L_s[U,\phid,\phi]+\hf L_s[U',\phi'^\dagger,\phi']
+[\sum\limits_{\v{m},\v{n}}\phid_j(\v{m})\Delta_{j,k}(\v{m},\v{n};U,U')\phi'_k(\v{n})+c.c]}
\ee
which must be a positive operator in the physical subspace of the Fock-space. This
condition will be expressed by the inequality
\be\label{reflpos}
\langle0|{\cal F}\Theta[{\cal F}]|0\rangle\ge0
\ee
holding for any local functional $\cal F$ including physical fields for positive $t$
only. Note that this condition obviously excludes states with negative norm from 
the physical subspace. 

The discussion of reversal parity in Section \ref{indnorm} makes clear that
the norm of a state created by an operator on a time reversal invariant vacuum
is determined by the time reversal parity of the operator in question. We shall use 
this relation to trace the physical states in the path integral of Eq. \eq{latpint}.
The time reversal acts as 
\be
\Theta{\cal F}[\phi,\phid,U]={\cal F}[\Theta\phi,\Theta\phid,\Theta U]
\ee
with 
\bea
\Theta\phi_j(n)&=&\phid_{\Theta j}(\Theta n),\nn
\Theta\phid_j(n)&=&\phi_{\Theta j}(\Theta n),\nn
\Theta U_{j,\mu}(n)&=&\begin{cases}U^\dagger_{\Theta j,\mu}(\Theta n)&\mu=1,2,3,\cr
U_{\Theta j-1,\mu}(\Theta n)&\mu=0,j<n_d,\cr
U_{j,\mu}(\Theta n-\hat0)&\mu=0,j=n_d,\end{cases}
\eea
where $\Theta j=n_d+1-j$ and the space-time coordinate $n=(t,\v{n})$ transforms as 
$\Theta(t,\v{n})=(-t,\v{n})$. 
We consider the site-inversion realization of time reversal, $t\to-t$, with odd $n_d$.
Had we chosen even $n_d$ we should have worked with the more involved
link-inversion, $t\to1-t$.
We shall need functionals with well defined time inversion parity,
$\Theta{\cal F}[\phi(n),\phid(n),U(n)]=\tau_{\cal F}{\cal F}[\phi(\Theta n),\phid(\Theta n),U(\Theta n)]$.
The simplest way to obtain such functionals is to use the combinations
\bea
\phi_{\tau,j}(t,\v{n})&=&\hf[\phi_j(t,\v{n})+\tau\phid_{\Theta j}(t,\v{n})],\nn
\phid_{\tau,j}(t,\v{n})&=&\hf[\phid_j(t,\v{n})+\tau\phi_{\Theta j}(t,\v{n})],\nn
U_{\tau,j,\mu}(t,\v{n})&=&\hf[U_{j,\mu}(t,\v{n})+\tau U^\dagger_{\Theta j,\mu}(t,\v{n})],~~~\mu=1,2,3
\eea
with $j=1,\ldots,(n_d-1)/2$ of the local fields which display well defined time reversal parities.
Note that the time reversal invariant combinations are the finite difference
realization of the coordinates $\phi_{\tau,j}$ mentioned above.
To achieve gauge invariance we perform the gauge transformation
\be\label{gaugef}
\omega(n_dt+j,\v{n})=[U_0(n_dt+j-1,\v{n})\cdots U_0(n_dt+1,\v{n})]^\dagger
\ee
for $2\le j\le n_d$ on the original lattice with simple time slices which cancels the 
time component of the gauge field within the block time slices and sets 
$U_{j,0}(t,\v{n})\to\openone$ for $j<n_d$. The functional $\cal F$ with time 
reversal parity $\tau_{\cal F}$ may contain any products of fields 
as long as the product of their time reversal parity agree $\tau_{\cal F}$.

The proof of the inequality \eq{reflpos} proceeds in the usual manner \cite{montvay},
by the splitting of the action \eq{blaction} into three pieces $S=L_s(0)+S_-+S_+$ with
\bea
S_-&=&\sum_{t<0}[L_s(t)+L_{km}(t)+L_{kg}(t)],\nn
S_+&=&\sum_{t\ge0}[L_{km}(t)+L_{kg}(t)]+\sum_{t>0}L_s(t)
\eea
We can now phrase an important condition of consistency, the time reversal invariance of the
dynamics of our model. The time reversal invariance of the microscopic dynamics
can be expressed by the equations $S_\pm[\Psi(t)]=\Theta[S_\mp[\Psi(t)]]=S_\mp[\Psi(\Theta t)]$
and $S_0[\Psi]=S_0[\Psi(\Theta t)]$. Furthermore the vacuum of the theory may 
be nontrivial but is assumed to contain time reversal invariant condensates only.

We introduce the notation $\Psi=(U,\phid,\phi)$ for the fields and write 
\be\label{prrp}
\langle0|{\cal F}\Theta {\cal F}|0\rangle=\int D[\Psi]e^{-S_0[\Psi]}e^{-S_+[\Psi]}{\cal F}[\Psi]e^{-S_-[\Psi]}\Theta[{\cal F}[\Psi]]
\ee
where the logarithm of the wave functional of the vacuum state is added to the 
actions $S_\pm[\Psi]$. To write this expression as an integral with positive definite
integrands we use the time reversal invariance of three objects. 
(i) The time reversal acts in a nontrivial manner on the wave functionals of the 
states, it flips the sign of the variables with negative time reversal parity. 
The time reversal invariance of the vacuum state, $\Theta|0\ra=|0\ra$, 
allows us to include the vacuum state functional into the time reversed factor
within the path integral,
\be
\langle0|{\cal F}\Theta {\cal F}|0\rangle=\int D[\Psi]e^{-S_0[\Psi]}e^{-S_+[\Psi]}{\cal F}[\Psi]\Theta\left[e^{-S_+[\Psi]}{\cal F}[\Psi]\right].
\ee
(ii) The time reversal invariance of $S_0[\Psi]$  is used to rewrite our expectation value as
\be
\langle0|{\cal F}\Theta {\cal F}|0\rangle=\int D_{t=0}[\Psi]
\int D_{t>0}[\Psi(t)]e^{-\hf S_0[\Psi(t)]-S_+[\Psi(t)]}{\cal F}[\Psi(t)]\Theta{\cal F}[\Psi(t)]
\int D_{\Theta t>0}[\Psi(\Theta t)]e^{-\hf S_0[\Psi(\Theta t)]-S_+[\Psi(\Theta t)]}
\ee
(iii) Finally, we assume that the functional ${\cal F}[\Psi]$ has time-reversal parity 
$\tau_{\cal F}$ and find
\be
\langle0|{\cal F}\Theta {\cal F}|0\rangle=\tau_{\cal F}\int D_{t=0}[\Psi]
\left(\int D_{t>0}[\Psi]e^{-\hf S_0[\Psi]-S_+[\Psi]}{\cal F}[\Psi]\right)^2
\ee
which is positive for $\tau_{\cal F}=1$. 

The naively expected conditions for consistency, the time reversal invariance
of the action and vacuum state are not enough for reflection positivity, we needed
condition (ii) satisfied by each trajectory in the path integral. The simplest local
way to ensure this condition together with (i) is to impose the boundary conditions 
\be
\Psi(t_f)=\tau_\Psi\Psi(t_i),
\ee
ie. using periodic or antiperiodic boundary conditions for time reversal even or odd
variables, respectively. Such a generalized KMS construction restricts the path integration
in a local, translationally invariant manner and eliminates the non-unitary runaway solutions. Full Poincar\'e invariance 
is recovered by imposing similar boundary conditions in each of space-time directions and performing
the thermodynamical limit.
Reflection positivity is assured within the Fock-space span by the local, 
time reversal invariant functionals of the fields acting on the time reversal invariant
vacuum assuming the generalized KMS conditions.

Due to the compactness of the proof it is instructive to check the result in a 
simple particular case of a free real scalar theory defined by the Lagrangian
$L=\phi D^{-1}(\Box)\phi/2$ and for the operators ${\cal F}=\partial_0^k\phi(x)$.
The space of field configurations to integrate over is determined by the 
generalized KMS boundary conditions $\phi^{(n)}_j(t_f,\v{x})=(-1)^j\phi^{(n)}_j(t_i,\v{x})$,
imposed on the original field with $j=0$ and the auxiliary fields for $j>0$.
The Lagrangian is quadratic in the auxiliary fields even in the presence of interactions
and the equations of motion can be used to eliminate these variables. The result is a path
integral for the original field satisfying the boundary conditions
\be\label{bcof}
\partial_0^j\phi(t_f,\v{x})=(-1)^j\partial_0^j\phi(t_i,\v{x}).
\ee
Eq. \eq{prrp} can now be written as
\be\label{freecheck}
\partial_t^k\partial_{t'}^k\la\phi(t,\v{x})\phi(t',\v{x})\ra_{|t'=\Theta t}
=\sum_n\partial_t^j\phi^{(n)}(t,\v{x})\partial_{t'}^j\phi^{(n)}(t',\v{x})_{|t'=\Theta t}\lambda^{(n)}
\ee
where the eigenfunction $D(\Box)\phi^{(n)}=\lambda^{(n)}\phi^{(n)}$ satisfies the boundary
condition \eq{bcof}. Since $\partial_t^k\phi^{(n)}(t,\v{x})=(-1)^k\partial_{t'}^k\phi^{(n)}(t',\v{x})_{|t'=\Theta t}$
each contribution to the sum on the right hand side of Eq. \eq{freecheck} is positive 
for even $k$ as long as the imaginary time path integral is convergent, $\lambda^{(n)}>0$.

\section{Summary}\label{sum}
The elimination of some high energy particle modes poses interesting problems for the
resulting low energy effective theory. An approximation which is necessary to render 
these theories useful is the truncation of the gradient expansion of the effective action. 
Two problems, related to the truncation of the effective theory have been considered 
in this work, the question of identifying the physical states of the effective theory 
and the proof that the low energy modes follow consistent dynamics.

The truncation of the gradient expansion of the effective dynamics generates
new degrees of freedom which display a generalization of the KMS construction
and are represented by periodic or antiperiodic trajectories
in the path integral representation. It was shown in the framework of the 
Yang-Mills-Higgs model with higher order derivatives that reflection positivity holds 
for this theory within the Fock-space generated by the action of local expressions of 
time reversal invariant operators on the time reversal invariant vacuum when these boundary
conditions are maintained. Note that such an extended boundary condition
prescription can always be imposed on any quantum field theory of the type considered in
this work because it modifies the propagators only when higher orders of the space-time 
derivatives occur in the action and the result holds for arbitrary value of the lattice spacing.

Further extensions of the argument presented above are needed to cover 
a bigger, more realistic family of effective theories. The higher order derivatives 
might appear in the terms containing higher order than quadratic of the matter 
fields. Furthermore, it is necessary to extend the argument for fermions
whose exchange statistics and lattice formalism introduces further complexities.

\end{document}